\documentclass[twocolumn,aps,floatfix,showpacs,superscriptaddress]{revtex4}
\usepackage{amsmath,amssymb,eucal,graphicx,bm}
\begin{document}
\title{Persistence of Random Walk Records}
\author{E.~Ben-Naim}
%\email{ebn@lanl.gov}
\affiliation{Theoretical Division and Center for Nonlinear
Studies, Los Alamos National Laboratory, Los Alamos, New Mexico
87545, USA}
\author{P.~L.~Krapivsky}
\affiliation{Department of Physics,
Boston University, Boston, Massachusetts 02215, USA}
%\email{paulk@bu.edu}
\begin{abstract}
  We study records generated by Brownian particles in one
  dimension. Specifically, we investigate an ordinary random walk and
  define the record as the maximal position of the walk.  We compare
  the record of an individual random walk with the mean record,
  obtained as an average over infinitely many realizations. We term
  the walk ``superior'' if the record is always above average, and
  conversely, the walk is said to be ``inferior'' if the record is
  always below average. We find that the fraction of superior walks,
  $S$, decays algebraically with time, $S\sim t^{-\beta}$, in the
  limit $t\to\infty$, and that the persistence exponent is nontrivial,
  $\beta=0.382258\ldots$.  The fraction of inferior walks, $I$, also
  decays as a power law, $I\sim t^{-\alpha}$, but the persistence
  exponent is smaller, $\alpha=0.241608\ldots$. Both exponents are
  roots of transcendental equations involving the parabolic cylinder
  function. To obtain these theoretical results, we analyze the joint
  density of superior walks with given record and position, while for
  inferior walks it suffices to study the density as function of
  position.
\end{abstract}
%05.40.Fb 	Random walks and Levy flights
%05.40.Jc 	Brownian motion
%02.50.Cw 	Probability theory
%02.50.Ey 	Stochastic processes
\maketitle

\section{Introduction}

The record, defined as the extremum in a sequence of variables, is a
useful characteristic of a dataset. Extreme value theory
\cite{wf,llr,eig,rse,sir} and analysis of sequences of uncorrelated
random variables \cite{abn,vbn} provide the basis for understanding
record statistics. Records in problems ranging from finance
\cite{ekm,bp,syn,wbk} and sport \cite{gts} to random structures
\cite{bkm,bk04} and complex networks \cite{kr02,gl,ggl} typically
involve sequences of {\em correlated} random variables. However,
current theoretical understanding of extreme values of correlated
random variables is still far from complete \cite{mk,jk,gw}.

Brownian trajectories are prime examples of correlated time series
\cite{pl,im,mp}.  Established record statistics of discrete time
one-dimensional random walks include the distribution of the number of
records and the mean duration of the longest record
\cite{sa,mz}. Records in ensembles of random walks, especially the
distribution of the maximum, have also been studied both for
independent \cite{kmr,wms} and for interacting random walks
\cite{smcrf,kik}. However, much less is known about random walk
records in higher dimensions \cite{ekb11,sm}.
 
Recent studies show that first-passage \cite{sr} and persistence
properties \cite{bms,dph} of records have rich phenomenology
\cite{bk13,mb}. For a sequence of uncorrelated random variables, the
probability that all records are above average decays algebraically
with sequence length and this behavior is governed by a nontrivial
persistence exponent. Such persistence characteristics were used to
analyze earthquake data \cite{bk13,mb}. In this article, we study
similar persistence characteristics of random walk records.

We consider a discrete time random walk in one dimension. The walk
starts at the origin, $x(0)=0$, and in each time step the walk makes a
jump: $x(t+1)=x(t)+\Delta_t$. The jump lengths $\Delta_t$ are
independent random variables chosen from a symmetric distribution with
finite variance: $\langle \Delta\rangle =0$ and $\langle\Delta_i
\Delta_j\rangle =\delta_{ij}\,\langle \Delta^2\rangle$ with $\langle
\Delta^2\rangle < \infty $.
 
The record $r(t)$ is defined as the maximal position of the random walk 
in the time interval $(0,t)$
\begin{equation} 
\label{rt-def}
r(t)=\text{max}\{x(0),x(1),x(2),\ldots,x(t)\}\,.
\end{equation}
We compare the record with the average record \hbox{$a(t)=\langle
  r(t)\rangle$} where the brackets denote an average over all possible
realizations of the random process governing the position $x(t)$.
Specifically, we compare the sequence of records
$\{r(0),r(1),r(2),\ldots,r(t)\}$ generated by the random walk with the
sequence of average records $\{a(0),a(1),a(2),\ldots,a(t)\}$. We call
a random walk {\it superior} if all records exceed the average,
$r(\tau)\geq a(\tau)$ for all $\tau=0,1,2,\ldots,t$. Similarly, we define an
{\it inferior} walk as one for which all records trail the average,
$r(\tau)\leq a(\tau)$ for all $\tau=0,1,2,\ldots,t$.

We now define $S(t)$ and $I(t)$ as the probability that at time $t$, a
walk is superior or inferior.  Our main result is that the
probabilities $S(t)$ and $I(t)$ decay algebraically with time
\begin{equation}
\label{SI-decay}
S\sim t^{-\beta} \qquad\text{and}\qquad I\sim t^{-\alpha}
\end{equation}
as $t\to\infty$. The persistence exponents are transcendental numbers
$\beta=0.382258\ldots$ and $\alpha=0.241608\ldots$.  Both exponents
are related to roots of the parabolic cylinder function,
\begin{equation}
\label{ab-root}
D_{2\beta+1}\left(\sqrt{2/\pi}\,\right)=0 \quad \text{and} \quad 
D_{2\alpha}\left(-\sqrt{2/\pi}\,\right)=0.
\end{equation}
The asymptotic behaviors \eqref{SI-decay}--\eqref{ab-root} apply as
long as the jump length distribution has zero mean and finite
variance. Hence, we can restrict our attention to random walks with
unit jump length: $\Delta=1$ and $\Delta=-1$ are chosen with equal
probabilities.

The rest of this paper is organized as follows. In Section II, we
briefly summarize basic properties of the record including the
distribution of records and the mean record. Since record is coupled
to position, we study the joint density of superior walks with given
record and position. This distribution obeys the diffusion equation,
and we obtain the long time asymptotic behavior using scaling analysis
(Section III). In the complementary case of inferior walks the
analysis simplifies because it suffices to consider only the position
(Section IV). A few generalizations are mentioned in Section V, and
concluding remarks are given in Section VI.

\section{The Average Record} 

We use a simple random walk as a model for Brownian motion in one
dimension \cite{mp,krb}.  The random walk starts at the origin, $x=0$
at time $t=0$, and in each time step, its position changes by a fixed
amount
\begin{equation}
\label{rw}
x(t+1) = 
\begin{cases}
x(t)-1 & \text{with prob.\ }1/2;\\
x(t)+1 & \text{with prob.\ }1/2.
\end{cases}
\end{equation}
With these jump rules, the average position does not change, $\langle
x(t)\rangle=0$, while the the mean square displacement equals time
$\langle x^2(t)\rangle = t\,$.

The record, defined in Eq.~\eqref{rt-def}, equals the maximum position
to date. For a simple random walk, the average record grows as the
square root of time
\begin{equation}
\label{at}
a(t)\simeq A\sqrt{t}\,, \quad\text{with}\quad A=\sqrt{2/\pi}\,.
\end{equation} 
This behavior represents the leading asymptotic behavior. Similar
behavior holds as long as the jump length distribution has zero mean
and a finite variance, and in general, the ratio between the average
record and the mean square displacement approaches a constant,
$a(t)/\sqrt{\langle x^2(t)\rangle}\to A$ in the limit $t\to\infty$
\cite{im,pl}.

Let $q(r,t)$ be the probability distribution function that the record
equals $r$ at time $t$. This quantity follows from the probability
that $r(\tau)<n$ for all $\tau=0,1,2,\ldots,t$.  The probability of this
event is the same as the survival probability $Q(n,t)$ that a random
walker starting at the origin never crosses $n$ during the time
interval $(0,t)$. The quantity $Q(n,t)$ is well known and can be
conveniently expressed using the error function, \hbox{$Q(n, t) =
  \text{erf}\left(n/\sqrt{2t}\,\right)$} in the long-time limit
\cite{sr,krb}.  The probability that $r(t)=n$ is then $Q(n+1,t)-Q(n,
t)$ which is asymptotically equivalent to $dQ(n,t)/dn$. As a result,
the probability distribution function $q(r,t)$ is a one-sided
Gaussian,
\begin{equation}
\label{qrt}
q(r,t)\simeq  \sqrt{\frac{2}{\pi t}}\,\exp\!\left(-\frac{r^2}{2t}\right) 
\end{equation}
for $r\geq 0$. In particular, we recover the probability that
the random walk remains in negative half-space, \hbox{$q(0,t)\simeq
A/\sqrt{t}$} \cite{sr}.

Let $M_n=\int_0^\infty dr\, r^n\, q(r,t)$ be the $n$th moment of the
record distribution. The zeroth moment, $M_0=1$, reflects that the
probability distribution is normalized, and the first moment,
$M_1\simeq A\sqrt{t}$, gives the average quoted in \eqref{at}.  The
second moment $M_2\simeq t$ gives the variance,
\begin{equation}
\label{var}
\langle r^2\rangle -\langle r\rangle^2=\Big(1-\frac{2}{\pi}\Big) t. 
\end{equation}
The average \eqref{at}, the variance \eqref{var}, and the distribution
function \eqref{qrt} show that the record grows as square-root of
time, $r\sim \sqrt{t}$. Hence, the typical record mimics the behavior
of the typical position $x\sim \sqrt{t}$.

\section{Superior Walks}

\begin{figure}[b]
\includegraphics[width=0.35\textwidth]{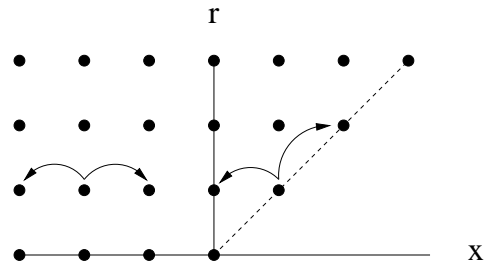}
\caption{Illustration of the jump processes \eqref{rw1} and \eqref{rw2}.}
\label{fig-ill}
\end{figure}

We now focus on superior walks, that is, walks for which the record
exceeds the average record, $r(\tau)\geq a(\tau)$ at all times
$\tau\leq t$. Since record $r$ is coupled to position $x$, we have to
consider how the pair of coordinates $(x,r)$ evolves with time. The
position changes at each time step. However, the record may or may not
change, and there are two possibilities.  When $x<r$ the position
changes but the record stays the same (see Fig.~\ref{fig-ill})
\begin{equation}
\label{rw1}
(x,r)\to 
\begin{cases}
(x-1,r) & \text{with prob.\ }1/2;\\
(x+1,r) & \text{with prob.\ }1/2.
\end{cases}
\end{equation}
When $x=r$, the position changes and depending on jump direction,
the record may increase,
\begin{equation}
\label{rw2}
(r,r)\to 
\begin{cases}
(r-1,r)   & \text{with prob.\ }1/2;\\
(r+1,r+1) & \text{with prob.\ }1/2.
\end{cases}
\end{equation}
As illustrated in Fig.~\ref{fig-ill}, the position performs ordinary
random walk in the $x-r$ plane, and there is also upward ``slip''
along the diagonal $x=r$. 

Let $P(x,r,t)$ be the density of superior walks with position $x$ and
record $r\geq x$ at time $t$.  A sum over all records and positions
gives the probability that a random walk is superior
\begin{equation}
\label{S-def}
S(t)=\sum_{r\geq 0}\,\sum_{x\leq r} P(x,r,t).
\end{equation}
In the discrete time formulation the sums are actually finite, $0\leq
r\leq t$ and $-t\leq x\leq r$.

The jump rules \eqref{rw1}--\eqref{rw2} imply the following recurrence
equations which relate the density at time $t+1$ to the density at
time $t$,
\begin{subequations}
\begin{align}
P(x,r,t+1) &=\frac{P(x\!-\!1,r,t)\!+\!P(x\!+\!1,r,t)}{2}
\label{Pxr}\\
P(r,r,t+1) &=\frac{P(r\!-\!1,r,t)\!+\!P(r\!-\!1,r\!-\!1,t)}{2}.
\label{Prr}
\end{align}
\end{subequations}
Equation \eqref{Pxr} is valid for all $x<r$ and it corresponds to
cases where the record was set before the final step.  Equation
\eqref{Prr} describes the evolution of the density along the diagonal
$x=r$ and it contains a contribution from walks in which the record
was set at the final step.

For the random walk \eqref{rw}, position $x$ and record $r$ are
discrete variables. Since we are interested in the long-time
asymptotic behavior, we may treat these variables as continuous.  The
density $P(x,r,t)$ satisfies the diffusion equation
\begin{equation}
\label{Pxr-eq}
\frac{\partial P}{\partial t}  = \frac{1}{2}\,\frac{\partial^2 P}{\partial x^2}
\end{equation}
in the domain $-\infty<x<r$ and $r>0$. This equation follows from the
recurrence equation \eqref{Pxr}, and it reflects that the position
undergoes an ordinary diffusion process. To obtain \eqref{Pxr-eq}, we
replace the left-hand side in \eqref{Pxr} with a first-order Taylor
expansion in time, $P+\partial P/\partial t$, and similarly, replace
the right-hand-side with a second order expansion in position
$x$. The diffusion equation \eqref{Pxr-eq} is subject to
the boundary condition
\begin{equation}
\label{bc}
2\,\frac{\partial P}{\partial x} + \frac{\partial P}{\partial r}= 0
\end{equation}
on the diagonal $x=r$. This relation, which properly accounts for the
 upward slip along the boundary, can be derived from the recurrence
 equation \eqref{Prr} by repeating the steps leading to \eqref{Pxr-eq}.
 
We now introduce a new variable $y$ which is a linear combination of
record and position
\begin{equation}
\label{y-def}
y=2r-x.
\end{equation}
With this transformation of variables $(r,x) \to (r, y)$, the
diffusion process takes place in the domain $y\geq r\geq 0$, and
importantly, the boundary condition \eqref{bc} simplifies to $\partial
P/\partial r= 0$, along the diagonal $y=r$. According to equation
\eqref{Pxr-eq}, the density $P\equiv P(y,r,t)$ still obeys the
diffusion equation
\begin{equation}
\label{Py-eq}
\frac{\partial P}{\partial t} = \frac{1}{2}\,\frac{\partial^2 P}{\partial y^2} 
\end{equation}
in the domain $y>r\geq 0$.   

As discussed in Section II, the position and the record both grow as 
square root of time, \hbox{$x\sim \sqrt{t}$} and
\hbox{$r\sim\sqrt{t}$}, and consequently, \hbox{$y\sim
 \sqrt{t}$}. Hence, the density of superior walks has the scaling form
\begin{equation}
\label{P-scaling}
P(y,r,t) \sim
t^{-\beta-1}\Phi\left(\frac{y}{\sqrt{t}}\,,\,\frac{r}{\sqrt{t}}\right).
\end{equation}
This scaling form is compatible with \eqref{S-def} and the algebraic
decay $S\sim t^{-\beta}$.  The scaling function $\Phi\equiv \Phi(Y,R)$
depends on the variables $Y=y/\sqrt{t}$ and $R=r/\sqrt{t}$
corresponding to the scaled position and record, respectively. For
superior walks we have $y\geq r\geq a$ and from equation \eqref{at},
we conclude $Y>R>A$ with $A=\sqrt{2/\pi}$. Hence, we have the boundary
condition $\Phi(Y,R)=0$ on the line $R=A$.

The boundary condition $\partial P/\partial r=0$ on the diagonal $y=r$
implies $\partial\Phi/\partial R=0$ when $Y=R$. This suggests to seek
a scaling function that depends on the variable $Y$ alone,
\hbox{$\Phi\equiv \Phi(Y)$}. By substituting the scaling form
\eqref{P-scaling} into the diffusion equation \eqref{Py-eq}, we find
that $\Phi$ obeys the second-order ordinary differential equation
\begin{equation}
\label{P-scaling-eq}
\Phi'' + Y\Phi'+2(\beta+1)\Phi=0,
\end{equation}
where prime denotes differentiation with respect to $Y$.  The
boundary condition is $\Phi(A)=0$. The first two terms in
\eqref{P-scaling-eq} imply that $\Phi$ has a Gaussian tail, \hbox{$\Phi\sim
\exp(-Y^2/2)$} when $Y\to\infty$. Next, we make the transformation
\hbox{$\Phi(Y)=\phi(Y)\exp\big(-Y^2/4\big)$} and arrive at the 
parabolic cylinder equation with index $2\beta+1$ \cite{bo} 
\begin{equation}
\phi'' +\left(2\beta + \frac{3}{2}-\frac{Y^2}{4}\right)\phi=0.
\end{equation}
This equation has two independent solutions: $D_{2\beta+1}(Y)$ and
$D_{2\beta+1}(-Y)$ where $D_\nu$ is the parabolic cylinder function of
index $\nu$. Since the density vanishes, \hbox{$\Phi\to 0$} as
$Y\to\infty$, we choose the former solution, and therefore, 
$\Phi(Y)=D_{2\beta+1}(Y)\exp(-Y^2/4)$.  The boundary condition
$\Phi(A)=0$ ``selects'' the persistence exponent $\beta$ as the
smallest root of the transcendental equation 
\begin{equation}
\label{beta}
D_{2\beta+1}(A)=0, 
\end{equation}
in agreement with the announced result \eqref{ab-root}. 

\begin{figure}[t]
\includegraphics[width=0.43\textwidth]{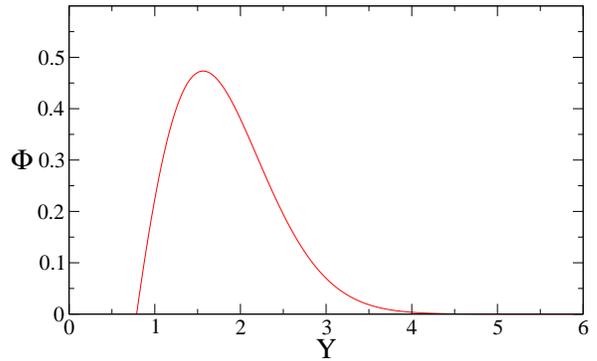}
\caption{The scaling function $\Phi(Y)$ versus the scaling variable
  $Y$.  The scaling function vanishes at $Y=A=\sqrt{2/\pi}$.}
\label{fig-phi}
\end{figure}

In terms of the original variables $x$ and $r$, the joint density
$P(x,r,t)$ has the asymptotic behavior (Fig.~\ref{fig-phi})
\begin{equation*}
P(x,r,t) \sim t^{-1-\beta}D_{2\beta+1}
\left(\frac{2r-x}{\sqrt{t}}\right)\exp\left[-\frac{(2r-x)^2}{4t}\right].
\end{equation*}
This distribution holds for $r>A\sqrt{t}$ and \hbox{$-\infty < x <r
  $}.  We obtained the density $P(x,r,t)$ up to a prefactor that
  cannot be determined using scaling analysis alone.  
Finally, the asymptotic behavior $D_\nu(z)\sim z^\nu \exp(-z^2/4)$
  shows that the density has a Gaussian tail $P(x,r,t)\sim
  \exp[-(2r-x)^2/(4t)]$.

It is straightforward to compute various moments of the joint
distribution. In the long-time limit, these moments are directly
related to moments of the scaling function $\Phi(Y)$, and it is
convenient to use the adjusted moments \hbox{$m_n=\int_A^\infty
dY\,(Y^n - A^n)\Phi(Y)$}.  Remarkably, the average position of
superior walks $\langle x\rangle_{\rm sup}$ coincides with the average
record, $\langle x\rangle_{\rm sup} \simeq A\sqrt{t}\,$. As expected,
the average record of superior walks grows faster, $\langle
r\rangle_{\rm sup} \simeq C\,\sqrt{t}$ with $C=m_2/(2m_1)$ or
$C=1.478591$. Finally, the position and the record are correlated
random variables, \hbox{$\langle xr\rangle_{\rm sup} \ne \langle
x\rangle_{\rm sup} \langle r\rangle_{\rm sup}$}, as follows from
\hbox{$\langle xr\rangle_{\rm sup}\simeq c\,t$} with $c=A^2/2+m_3/(6m_1)$.

\section{Inferior Walks}

Inferior walks are simpler to analyze because they can be defined in
terms of position alone: A walk is inferior if and only if
$x(\tau)\leq a(\tau)$ for all $\tau=0,1,2,\ldots,t$.  Indeed, if the
position exceeds the average record, the record necessarily crosses
the average.  Conversely, if the position never exceeds the average
record, then the record remains below average.  Hence, inferior walks
map onto diffusion in the presence of a receding trap with location
that grows as square root of time, a problem that was solved in
Ref.~\cite{kr}.

Since it is not necessary to keep track of the record, we study the
distribution of position. Let $P(x,t)$ be the density of inferior
walks with position $x$ at time $t$. The probability $I(t)$ that a
walk remains inferior after $t$ steps is the integral of the density,
\hbox{$I(t)=\int_{-\infty}^{a(t)} dx P(x,t)$}. The density of inferior
walks obeys the diffusion equation
\begin{equation}
\label{Pxt-eq}
\frac{\partial P(x,t)}{\partial t}  = 
\frac{1}{2}\frac{\partial^2 P(x,t)}{\partial x^2},
\end{equation}
in the domain $-\infty < x < a(t)$, and is subject to the boundary
condition $P(a,t)=0$. We anticipate the scaling behavior 
\begin{equation}
\label{Pxt-scaling}
P(x,t)\sim t^{-\alpha-1/2}\,\Psi\left(\frac{x}{\sqrt{t}}\right),
\end{equation}
and impose the boundary condition $\Psi(A)=0$. The prefactor in
\eqref{Pxt-scaling} reflects the algebraic decay $I(t)\sim
t^{-\alpha}$.

Substituting the scaling form \eqref{Pxt-scaling} into the diffusion
equation \eqref{Pxt-eq}, we find that the scaling function obeys 
\begin{equation}
\label{Psi-eq}
\Psi''+X\Psi'+(2\alpha+1)\Psi=0.
\end{equation}
Here, prime denotes differentiation with respect to the scaling
variable $X=x/\sqrt{t}$.  With the transformation
\hbox{$\Psi(X)=\psi(X) \exp(-X^2/4)$}, the function $\psi(X)$ obeys
the parabolic cylinder equation
\begin{equation}
\label{psi-eq}
\psi''+\left(2\alpha+\frac{1}{2}-\frac{X^2}{4}\right)\psi=0.
\end{equation}
This equation has two linearly independent solutions: $D_{2\alpha}(X)$
and $D_{2\alpha}(-X)$. The density should vanish as $X\to -\infty$
and this requirement gives $\psi(X)= D_{2\alpha}(-X)$.  The boundary
condition $\Psi(A)=0$ leads to the transcendental equation stated
in \eqref{ab-root},
\begin{equation}
\label{alpha}
D_{2\alpha}(-A)=0.
\end{equation}

In terms of the original variables, the density of inferior walks has
the asymptotic behavior
\begin{equation}
\label{Pxt}
P(x,t) \sim t^{-\alpha-1/2}\,D_{2\alpha}\!\left(-\,\frac{x}{\sqrt{t}}\right)
\exp\!\left[-\,\frac{x^2}{4t}\right].
\end{equation}
In particular, the density has a Gaussian tail \hbox{$P(x,t)\sim
\exp(-x^2/2t)$} as $x\to -\infty$. 

Thus far, we considered only the maximal position, but one can also
consider the maximal and minimal positions simultaneously.  When
$|x(\tau)|\leq a(\tau)$ for all \hbox{$\tau=0,1,2,\ldots,t$}, the
minimal position and the maximal position are both inferior with
respect to the average record.  The density of such {\em meek} random
walks satisfies the diffusion equation \eqref{Pxt-eq} in the growing
interval $\big[-a(t), a(t)\big]$. We seek a scaling solution
\hbox{$P(x,t)\sim t^{-\gamma-1/2}\,\Psi(X)$} and arrive at the same ordinary
differential equations \eqref{Psi-eq}--\eqref{psi-eq} with a new
persistence exponent $\gamma$ replacing $\alpha$.  The density is
symmetric, $\psi(X)=\psi(-X)$, and thus,
$\psi(X)=D_{2\gamma}(X)+D_{2\gamma}(-X)$. The boundary condition
$\Psi(A)=\Psi(-A)=0$ leads to the transcendental equation
\begin{equation}
\label{gamma}
D_{2\gamma}(A) + D_{2\gamma}(-A)=0
\end{equation}
that specifies the persistence exponent $\gamma=1.698282\ldots$.  The
probability that a walk is meek, $\int_{-a}^a dx P(x,t)\sim
t^{-\gamma}$, is therefore much smaller than the probability that the
walk is either superior or inferior.

\section{Simulations and Extensions} 

\begin{figure}[t]
\includegraphics[width=0.43\textwidth]{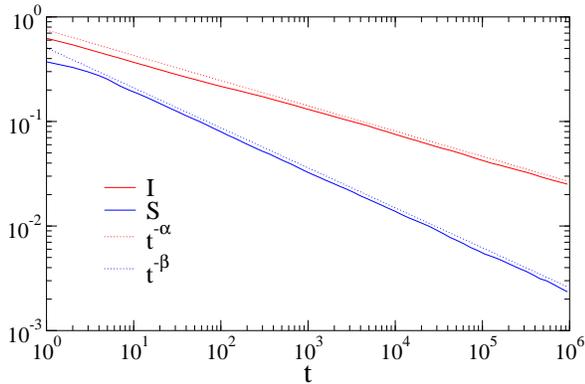}
\caption{The fractions $S$ and $I$ of superior and inferior walks
  versus time $t$. The simulation results represents $10^6$
  independent Monte Carlo runs. Also shown for reference are the
  theoretical results \eqref{SI-decay}--\eqref{ab-root}.}
\label{fig-st}
\end{figure}

Figure \ref{fig-st} shows results of Monte Carlo simulations for the
fractions $S$ and $I$ of superior and inferior walks. These results
are in excellent agreement with the theoretical predictions. In the
computations, we first calculated the average record $a(t)$ by
considering $M$ independent realizations of a discrete-time random
walk with $t$ steps. We then measured the fraction of all of
realizations in which the maximal position of the walk is always
larger or always smaller than the average record, $r(\tau)\geq
a(\tau)$ or $r(\tau)\leq a(\tau)$ for all $\tau\leq t$.

To verify that the asymptotic behavior
\eqref{SI-decay}--\eqref{ab-root} is robust, we considered two
distributions of jump length: (i) fixed step size: $\Delta=1$ or
$\Delta=-1$ with equal probabilities, and (ii) a variable step size,
chosen with uniform probability in the domain $-1\leq \Delta \leq
1$. The results shown in figure \ref{fig-st} are for the latter case
and were obtained using $M=10^6$ independent realizations.

\begin{figure}[t]
\includegraphics[width=0.43\textwidth]{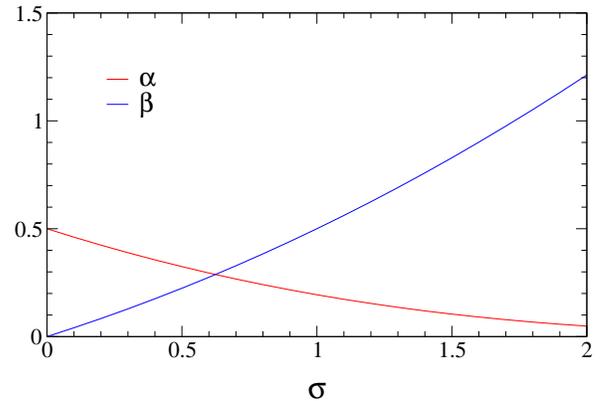}
\caption{The exponents $\alpha$ and $\beta$ characterizing $\sigma-$inferior 
and $\sigma-$superior random walks.}
\label{fig-beta}
\end{figure}

Our analysis compared the record $r$ with the average record $a\simeq
A\sqrt{t}$. We can compare the record with other length scales that
grow as square root of time. We thus define a walk to be
$\sigma-$superior [respectively $\sigma-$inferior] if $r(\tau)\geq
\sigma\,\sqrt{\tau}$ [respectively $r(\tau)\leq \sigma\,\sqrt{\tau}$]
for all $\tau=0,1,2,\ldots,t$. A straightforward generalization of
the above analysis shows that the persistence exponents
$\alpha(\sigma)$ and $\beta(\sigma)$ that govern the abundance of
such walks are given by (Figure \ref{fig-beta})
\begin{equation}
\label{ba}
D_{2\alpha}(-\sigma)=0 \quad\text{and}\quad D_{2\beta+1}(\sigma)=0.
\end{equation}
As expected, the persistence exponent \hbox{$0<\alpha<\infty$} increases
monotonically with $\sigma$, while the exponent \hbox{$0<\beta<1/2$}
decreases monotonically. The maximal value $\alpha(0)=1/2$ follows
from the probability that a random walk remains in the negative half
space, \hbox{$S\sim t^{-1/2}$} \cite{sr}.  The exponent $\alpha$
decays rapidly, \hbox{$\alpha\simeq \left(\sigma/\sqrt{8\pi}\right)
e^{-\sigma^2/2}$}, while the exponent $\beta$ diverges,
\hbox{$\beta\simeq \sigma^2/8$}, in the limit $\sigma\to \infty$ (see
\cite{bk10}).

Figure \ref{fig-beta} shows that both $\alpha<\beta$ and
$\alpha>\beta$ are possible and that both persistence exponents vary
continuously with $\sigma$. We note that for uncorrelated random
variables, it was also found that $\alpha<\beta$ and $\alpha>\beta$
are both feasible, and that both exponents are continuous functions of
some control parameter \cite{bk13}. However, we do not believe that
there is a deeper connection between these two sets of exponents or
that it is possible to obtain persistence characteristics of records
by mapping correlated random variables onto uncorrelated random
variables.

Another useful by-product of our analysis is the joint distribution
$P(x,r,t)$ of position and record for a one-dimensional random walk.
If we do not impose any restriction on the record then $\beta=0$ in
Eqs.~\eqref{P-scaling}-\eqref{P-scaling-eq}. The corresponding
solution of \eqref{P-scaling-eq} with $\beta=0$ is
$D_1(Y)\exp(-Y^2/4)$ and therefore,
\begin{equation}
\label{Pxrt}
P(x,r,t) \simeq \sqrt{\frac{2}{\pi t^3}}
\,(2r-x)\,\exp\left[-\frac{(2r-x)^2}{2t}\right].
\end{equation}
Normalization of the probability distribution sets the numerical
prefactor. By integrating \eqref{Pxrt} over position, one recovers the
record distribution \eqref{qrt}.  The joint distribution \eqref{Pxrt}
has been originally discovered by L\'evy, see \cite{pl,im}.  

The joint distribution \eqref{Pxrt} shows that record and position are
correlated variables. In particular, $\langle xr\rangle \simeq t/2$
whereas $\langle x\rangle=0$ and $\langle r\rangle \simeq
A\sqrt{t}$. Higher-order moments also reflect that $x$ and $r$ are
correlated, and for example, $\langle x^2r^2\rangle \simeq 2t^2$
whereas $\langle x^2\rangle = \langle r^2\rangle = t$.

\section{Conclusions}

In conclusion, we used the average record to characterize the motion 
of a Brownian particle in one dimension. A random walk is said to be
superior if its maximal position is always above average and
similarly, it is said to be inferior if the maximal position is always
below average. We find that the probability that a walk is superior or
inferior decays algebraically with time. This power-law decay is
characterized by nontrivial persistence exponents.

For inferior walks, it suffices to keep track of the position of the
walk alone and consequently, the problem reduces to diffusion in the
presence of a properly-chosen moving trap. For superior walks, it is
necessary to keep track of both record and position, and consequently,
the problem involves diffusion in two-dimensional space. This random
process consists of diffusion in the position coordinate and
directional motion when the record and the position are
equal. Nevertheless, a linear transformation of variables reduces the
problem to one-dimensional diffusion in the presence of a moving trap
as far as the asymptotic behavior is concerned (Figure \ref{fig-ill}).

Our results address the leading asymptotic behavior. However,
different implementations of a random walk are not entirely 
equivalent.  For instance, there are corrections to the leading
behavior of the average \eqref{at}, and $a(t)/\sqrt{\langle
x^2(t)\rangle} = A + C\,t^{-1/2} + \mathcal{O}(t^{-1})$. The constant
$C$ depends on the distribution of jump lengths and its derivation
requires rather intricate analysis \cite{cffh,cm,ss}. It will be
interesting to understand how corrections to the leading asymptotic
behavior \eqref{SI-decay}--\eqref{ab-root} depend on the distribution
of jump lengths.

Finally, we mention that records have been extensively used to analyze
empirical data such as earthquake inter-event times \cite{nmt,sdt} and
temperature readings \cite{ekbhs,rp}. Comparing a sequence of records
with a baseline such as the average provides a measure of performance.
The fraction of superior and inferior record sequences has shown to be
a sensible tool for analyzing earthquake data
\cite{bk13,mb}. Persistence of records is also useful in finance where
it is natural to compare an individual stock price with the stock
index \cite{ekm,bp,syn}.
 
\medskip

We thank Satya Majumdar for useful discussions and correspondence and
acknowledge DOE grant DE-AC52-06NA25396 for support (EB).


\begin{thebibliography}{99}

\bibitem{wf} 
      W.~Feller, 
      {\it An Introduction to Probability Theory and Its Applications}
      (Wiley, New York, 1968).

\bibitem{llr}
       M.~R.~Leadbetter, G.~Lindgren, H.~Rootzen, 
       {\it Extremes and Related Properties of Random Sequences and Processes},
        (Springer, Berlin, 1983).

\bibitem{eig}
       E.~I.~Gumbel, 
       {\it Statistics of Extremes} 
       (Dover, New York 2004).

\bibitem{rse}
       R.~S.~Ellis, 
       {\it Entropy, Large Deviations, and Statistical Mechanics}
       (Springer, Berlin 2005).

\bibitem{sir} 
       S.~I.~Resnick, 
       {\it Extreme Values, Regular Variation and Point Processes} 
       (Springer, Berlin, 2007).

\bibitem{abn}
       B.~C.~Arnold, N.~Balakrishnan and H.~N.~Nagraja, 
       {\it Records} (Wiley-Interscience, 1998). 
     
\bibitem{vbn}
       V.~B.~Nevzorov, {\em Records: Mathematical Theory},  
       Translation of Mathematical Monographs {\bf 194} 
       (American Mathematical Society, Providence, RI, 2001). 

\bibitem{ekm}
        P.~Embrechts, G.~Kl\"{u}ppelberg and T.~Mikosch,
        {\it Modelling extremal events for insurance and finance} 
        (Springer-Verlag, Berlin, 1997).

\bibitem{bp}
      J.~-P.~Bouchaud and M.~Potters, 
      {\em Theory of Financial Risk and Derivative Pricing}
      (Cambridge University Press, Cambridge 2003).

\bibitem{syn}
      S.~Y.~Novak,  
      {\it Extreme value methods with applications to finance}
      (Chapman \& Hall/CRC Press, London, 2011). 

\bibitem{wbk}
      G.~Wergen, M.~Bogner, and J.~Krug, 
      Phys.\ Rev.\ E {\bf 83}, 051109 (2011).

\bibitem{gts}
      D.~Gembris, J.~G.~Taylor, and D.~Suter, 
      Nature {\bf 417}, 506 (2002);
      J. Appl. Stat. {\bf 34}, 529 (2007).

\bibitem{bkm}  
      E.~Ben-Naim, P.~L.~Krapivsky,  and S.~N.~Majumdar, 
      Phys.\ Rev. E {\bf 64}, R035101 (2001). 

\bibitem{bk04}  
      E.~Ben-Naim and P.~L.~Krapivsky,  
      Europhys.\ Lett. {\bf 65}, 151 (2004).

\bibitem{kr02}  
      P.~L.~Krapivsky and S.~Redner, 
      Phys.\ Rev.\ Lett. {\bf 89}, 258703 (2002).

\bibitem{gl}
     C.~Godreche and J.~M.~Luck, 
     J. Stat. Mech.  P11006 (2008). 
       
\bibitem{ggl}
     C.~Godreche,  H.~Grandclaude, and J.~M.~Luck, 
     J. Stat. Mech.  P02001 (2010). 
       
\bibitem{mk}  
       S.~N.~Majumdar and P.~L.~Krapivsky, 
       Physica A {\bf 318}, 161 (2003). 

\bibitem{jk}
      J.~Krug,
      J. Stat. Mech. P07001 (2007).

\bibitem{gw}
      G.~Wergen,
      J. Phys. A {\bf 46}, 223001 (2013).
       
\bibitem{pl}
      P.~L\'evy, 
      {\it Processus Stochastiques et Mouvement Brownien} 
      (Gauthier-Villars, Paris, 1948) 

\bibitem{im} 
      K.~It\^{o} and H.~P.~McKean, 
      {\it Diffusion Processes and Their Sample Paths} 
      (New York, Springer, 1965).

\bibitem{mp} 
      P. M\"orders and Y.~Peres, 
      {\it Brownian Motion}
      (Cambridge University Press, Cambridge, 2010).     

\bibitem{sa}
       E.~Sparre Andersen, 
       Math. Scand. {\bf 1}, 263 (1953); {\it ibid}. {\bf 2}, 195 (1954). 

\bibitem{mz}
      S.~N.~Majumdar and R.~M.~Ziff,
      Phys. Rev. Lett. {\bf 101}, 050601 (2008).

\bibitem{kmr}  
      P.~L.~Krapivsky,  S.~N.~Majumdar, and A.~Rosso, 
      J.\ Phys.\ A {\bf 43}, 315001 (2010). 

\bibitem{wms}
      G.~Wergen, S.~N.~Majumdar, and G.~Scher, 
      Phys. Rev. E {\bf 86}, 011119 (2012).

\bibitem{smcrf}
      G.~Scher,  S.~N.~Majumdar, A. Comtet, and J.~Randon-Furling,       
      Phys. Rev. Lett. {\bf 101}, 150601 (2008). 
      
\bibitem{kik}      
      N.~Kobayashi, M.~Izumi, and M.~Katori,
      Phys.\ Rev.\ E {\bf 78}, 051102 (2008).

\bibitem{ekb11}
      Y.~Edery, A.~Kostinski, and B.~Borkowitz, 
      Geophys. Res. Lett. {\bf 389}, L16403 (2011). 

\bibitem{sm}
      S.~N.~Majumdar, Physica A {\bf 389}, 4299 (2010). 

\bibitem{sr} 
      S.~Redner, 
      {\it A Guide to First-Passage Processes} 
      (Cambridge University Press, Cambridge, 2001).

\bibitem{bms}
      A.~J.~Bray, S.~N.~Majumdar, and G.~Schehr, 
      {\em Persistence and First-Passage Properties in 
      Non-equilibrium Systems}, 
      Adv. in Phys., {\bf 62}, 225 (2013).

\bibitem{dph}
      B.~Derrida, V.~Hakim, and V.~Pasquier, 
      Phys. Rev. Lett. {\bf 75}, 751 (1995).

\bibitem{bk13}  
      E.~Ben-Naim and P.~L.~Krapivsky, Phys.\ Rev.\ E {\bf 88}, 022145 (2013).

\bibitem{mb}
      P.~W. Miller and E.~Ben-Naim, J. Stat. Mech. P10025 (2013). 

\bibitem{krb}  
       P.~L.~Krapivsky, S.~Redner and E.~Ben-Naim,
       {\it  A Kinetic View of Statistical Physics}
       (Cambridge University Press, Cambridge, UK, 2010).

\bibitem{bo}
      C.~M.~Bender and S.~A.~Orszag,
      {\em Advanced Mathematical Methods for Scientists and Engineers}
      (McGraw-Hill, New York, 1978).
                
\bibitem{kr}
      P.~L.~Krapivsky and S.~Redner, 
      Amer. Jour. Phys. {\bf 64}, 546 (1996).

\bibitem{bk10}
      E.~Ben-Naim and P.~L.~Krapivsky,  J. Phys. A {\bf 43}, 495007 (2010); 
      E.~Ben-Naim and P.~L.~Krapivsky,  J. Phys. A {\bf 43}, 495008 (2010).

\bibitem{cffh}
      E.~G.~Coffman, P.~Flajolet, L.~Flato, and M.~Hofri, 
      Probab. Eng. Inform. Sci. {\bf 12}, 373 (1998).

\bibitem{cm} 
      A.~Comtet and S.~N.~Majumdar, J. Stat. Mech. P06013 (2005).

\bibitem{ss}
        S.~Sabhapandit,
        EPL {\bf 94}, 20003 (2011).

\bibitem{nmt}
      W.~I.~Newman, B.~D.~Malamud, and D.~L.~Turcotte, 
      Phys. Rev. E {\bf 82}, 066111 (2010).

\bibitem{sdt} 
       R.~Shcherbakov, J.~Davidsen, and K.~F.~Tiampo, 
       Phys. Rev. E {\bf 87}, 052811 (2013).

\bibitem{ekbhs} 
      J.~F.~Eichner, E.~Koscielny-Bunde, A.~Bunde,
      S.~Havlin, and H.~J.~Schellnhuber,
      Phys. Rev. E {\bf 68}, 046133 (2003).

\bibitem{rp}
      S.~Redner and M.~R.~Petersen,
      Phys. Rev. E {\bf 74}, 061114 (2006).

\end{thebibliography}
\end{document}